\documentclass[12pt]{article}
\usepackage{pstricks}
\usepackage{pst-coil}
\usepackage{amssymb}
\begin{document}
\title{Sectors of Spherical Homothetic Collapse}
\author{Brien C. Nolan\\School of Mathematical Sciences,\\
Dublin City University,\\Glasnevin, Dublin 9\\Ireland}
\date{\today}
\maketitle
\begin{abstract}


\newcommand{\be}{\begin{eqnarray}}
\newcommand{\ee}{\end{eqnarray}}
\newtheorem{prop}{Proposition}

A study is undertaken of the gravitational collapse of spherically symmetric thick shells admitting a homothetic Killing vector field under the assumption that the energy momentum tensor corresponds to the absence of a pure outgoing component of field. The energy-momentum tensor is not specified beyond this, but is assumed to satisfy the strong and dominant energy conditions. The metric tensor depends on only one function of the similarity variable and the energy conditions identify a class of functions ${\cal F}$ to which the metric function may belong. The possible global structure of such space-times is determined, with particular attention being paid to singularities and their temporal nature (naked or censored). It is shown that there are open subsets of ${\cal F}$ which correspond to naked singularities; in this sense, such singularities are stable. Furthermore, it is shown that these singularities can arise from regular (continuous), asymptotically flat initial data which deviate from the trivial data by an arbitrarily small amount.
\end{abstract}

\section{Introduction and summary}
The most significant open problem of general relativity is that of cosmic censorship, encompassing as it does such a wide range of topics as black hole physics and the global initial value problem. There exist several different and sometimes inequivalent phrasings of this problem, but they all centre on the same issue: what is the relationship between the temporal nature of singularities and the initial data which give rise to them? The cosmic censorship hypothesis asserts that, roughly, physically realistic initial data cannot give rise to singularities which are visible to observers (see \cite{waldrev} for a more precise formulation). There are many caveats: the phrase `physically realistic' needs to be interpreted appropriately and should include some precise notion of being generic. For example, and significantly for the following, initial data which have zero measure in some general space of initial data sets and which lead to naked singularities should not be considered physically significant. Since no data can be measured with infinite precision, the {\em physical} initial data would overlap with the region of initial data space which leads to censored singularities. Additionally, the matter fields involved should obey a suitable energy condition and should be such that the Einstein-matter field equations have a well-posed initial value problem. In particular, the matter should be such that it evolves singularity-free (i.e. satisfies a well-posed global initial value problem) in Minkowski space-time. Otherwise any singularities which evolve may not be due to gravitational influences and hence do not relate to the question of cosmic censorship \cite{waldrev}.

There are many significant recent results which work in favour of the cosmic censorship hypothesis. In particular, the hypothesis (and much more besides) has been proven for the case of small data for various matter models \cite{ck,klain,rr,christo}. See \cite{waldrev} for a review of the status of the weak hypothesis (no globally naked singularities from generic, regular initial data). The main body of evidence in favour of the strong hypothesis (no locally naked singularities from generic, regular initial data) comes from studies of the stability of the Cauchy horizon in black hole interiors. See \cite{brady-kyoto} for an up to date review. Indirect, but equally significant evidence comes from the rigidity theorems for event and Cauchy horizons. See \cite{racz} for recent results and references.

There are of course a plethora of solutions of Einstein's equation which admit naked singularities. Those which have received most attention, and which can arise from regular initial data, are those generated in spherical dust collapse \cite{jj}, spherical massless scalar field collapse \cite{christo1} and self-similar spherical collapse \cite{ori-piran,lake,carr-coley}. All refer to highly symmetric specialised situations. Nonetheless, they contribute greatly to our understanding of the issues involved in cosmic censorship. Indeed the study of the singularities arising in massless scalar field collapse led to a weak cosmic censorship theorem for this case \cite{christo} which, among other things, may serve as a template for generalisations. Thus there may be much to be learned from studying such specialisations. 

In this paper we study cosmic censorship for such a specialised situation. We consider spherically symmetric self-similar space-times of a certain class (`no pure outgoing radiation'). See the following section for a complete description of the class of space-times studied. By self-similarity, we mean what Carr and Coley \cite{carr-coley} refer to self-similarity of the first kind; space-time $(M,g)$ admits a homothetic Killing vector field ${\vec \xi}$:
\[ {\cal L}_{{\vec \xi}}g_{ab}=2g_{ab}.\]
Using an advanced time coordinate $v$ and the radius function $r$ as coordinates in the radial 2-space, we can write the line element as
\[ ds^2 =-2Fe^{2\psi}dv^2+2e^{\psi}dvdr+r^2d\Omega^2,\]
where $F$ and $\psi$ depend only on the similarity variable $x=v/r$ and the homothetic Killing field is
\[ {\vec \xi}=v\frac{\partial}{\partial v}+r\frac{\partial}{\partial r}.\]

While assuming spherical symmetry and self-similarity is a severe 
restriction on the allowed structure of space-time, both these 
symmetries play an important role in gravitation. Spherical 
symmetry is a useful approximation for a collapsing compact 
object which becomes more accurate at the later stages of 
collapse as the object radiates away its higher multipole moments. 
The role of self-similarity in general relativity has recently been 
recognised as being quite significant, in particular as a symmetric state
which various systems approach asymptotically \cite{carr-coley} and 
in critical phenomenon \cite{gundlach}.

Numerous authors have studied spherically symmetric self-similar collapse
\cite{gold-pir,lake,henrik,ori-piran,brady}; see section 4.1 of \cite{carr-coley} 
for an up to date review. The authors all find that significant sectors of the classes of solutions being studied admit naked singularities; in some instances (e.g. \cite{ori-piran,brady}) these can arise from regular initial data which are asymptotically flat.

As yet, the stability of these solutions and their Cauchy horizons
has not been completely determined, and so their significance for cosmic censorship remains
in question. See however Section 4.4 of \cite{carr-coley} and \cite{waugh-lake} for results on stability; the latter have shown that the blue-sheet instability associated with the Cauchy horizon in black hole interiors is not present in self-similar collapse.
There is an associated stability issue which also needs to be addressed, namely
are the solutions admitting naked singularities stable in the class of solutions being studied?
If the answer is negative, then these solutions certainly do not provide significant evidence against
cosmic censorship. Of course one must also determine if the solutions arise from
regular, asymptoticlly flat initial data, i.e. do the solutions describe gravitational collapse
of a compact object?

We now give a brief description of what follows and a summary of our main results.
The physical model under discussion is described in Section 2. 
This can be described as the gravitational collapse of spherically symmetric, self-similar thick shells of matter/energy which do not admit any pure outgoing radiation. The inner boundary of the shell traces a spherical null hypersurface in space-time, interior to which space-time is flat. The matching problem across this boundary is addressed. The line element is shown to depend on just one function $F$ of the similarity variable $x=v/r$.
As stated, no particular matter model is assumed, but the matter is assumed to satisfy the strong and dominant energy conditions. The price paid for this assumption of a most general possible matter distribution is that we are leaving aside the issue of whether or not the Einstein-matter field equations are well posed. We will take account of this when drawing conclusions from this study. In particular, we will only be able to address the role of this hypothesis in the cosmic censorship conjecture, and not the conjecture itself. The energy conditions are discussed in Section 3 and it is shown how they identify in a natural manner a class of functions ${\cal F}$ to which $F$ may belong. This in turn identifies the class of space-times which are the subject of this paper. We point out the existence of singularities at the centre $r=0$ of these space-times.

The conformal diagrams for these space-times are given in Section 4. We classify the space-times according to (i) the asymptotic behaviour (as $x\to\infty$) of $F$, which relates to the occurence or otherwise of an apparent horizon, and (ii) the occurrence of naked singularities, which depends on the existence and number of intersections of the graphs of $F(x)$ and $1/x$. This classification is made possible by virtue of the structure of ${\cal F}$. There are nine inequivalent conformal diagrams (i.e. inequivalent classes); all but 3 of these admit naked singularities. One admits only locally naked singularities, all the others admit globally naked singularities.

In Section 5 we discuss the topological stability of the subclasses of solutions in the set ${\cal F}$. The energy conditions allow us to introduce a natural topology on ${\cal F}$. It is shown that the sets corresponding to globally naked singularities and to censored singularities are stable in the sense that they have open subsets. The set of locally naked singularities is shown to be unstable; it lies on the boundary between globally naked and censored singularities. An unstable subset of censored singularities is also identified.

The initial value problem for these space-times is discussed in Section 6. A regular initial data set is identified and an appropriate measure of the initial data is defined. We show that if the data are sufficiently small (i.e. sufficiently close to the trivial data, for which the evolution leads uniquely to Minkowski space-time), then the evolution, independent of the matter source apart from adherence to the energy conditions, necessarily leads to naked singulariites. We also give conditions on the initial data, essentially that they be sufficiently large, which guarantees that the evolution leads to a censored singularity. By introducing a cut-off point (or rather 2-sphere) on the initial data surface, beyond which the data are that of the Schwarzschild vacuum, the initial data are made to be asymptotically flat. However if the 3-metric and extrinsic curvature of this patched initial data surface are respectively differentiable and continuous, then the energy conditions are violated: all the solutions of this class are ruled out. Requiring that the 3-metric and extrinsic curvature be continuous allows all sectors of solutions found above. Thus we demonstrate the existence of asymptotically flat initial data, which are arbitrarily close to the trivial data and which allow the development of naked singularities. The data are continuous, but not differentiable. Some concluding comments are given in Section 7.

Our main results are given in the list of propositions in Sections 5 and 6. For brevity, the results of Section 4 are presented only in summary and in the conformal diagrams, Figures 2-10. We use the conventions of \cite{HE}, and a bullet $\bullet$ indicates the end (or absence) of a proof.

\section{The physical model}

We consider the following situation. Minkowski space-time is perturbed by the radial infall of gravitating field, which propagates inwards from ${\cal J}^-$. The influx is switched on at a certain finite time, and so the deviation from flatness commences on a spherical null hypersurface ${\cal N}$ of constant advanced time. Thus space-time consists of the union of ${\cal M}_1$, which is a portion of Minkowski space-time and lies to the past of ${\cal N}$, and a self-similar region ${\cal M}_2$, which lies to the future of ${\cal N}$. The situation is illustrated in Figure 1.

\begin{figure}
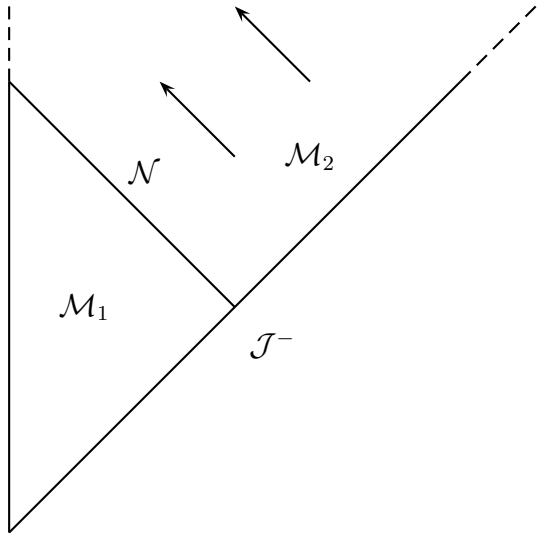

\pspicture*(-4,-4)(8,8)
$
\psline(0,0)(0,6)
\psline[linestyle=dashed](0,6)(0,7)
\psline(0,0)(6,6)
\psline[linestyle=dashed](6,6)(7,7)
\psline(3,3)(0,6)
\psline[arrows=->](3,5)(2,6)
\psline[arrows=->](4,6)(3,7)
\rput(3.5,2.5){\psframebox*[framearc=.3]{{\cal J}^-}}
\rput(1,3){\psframebox*[framearc=.3]{{\cal M}_1}}
\rput(1.8,4.8){\psframebox*[framearc=.3]{{\cal N}}}
\rput(4,5){\psframebox*[framearc=.3]{{\cal M}_2}}
$
\endpspicture
\caption{Conformal diagram of the physical model. ${\cal M}_1$ is a portion of Minkowski space-time; the region ${\cal M}_2$ is 
self-similar and contains no pure outgoing field. The arrows represent the influx from past null infinity ${\cal J}^-$. The spherical null hypersurface ${\cal N}$ divides the two regions of space-time.}
\end{figure}

Taking $V$ to be advanced time in ${\cal M}_1$, we can write the line element as
\[ ds^2 =-dV^2+2dVdr + r^2d\Omega^2,\]
where $d\Omega^2$ is the line element for the unit 2-sphere, and taking $v$ to be an advanced time coordinate in ${\cal M}_2$, we can write the line element of this region in the form
\begin{equation}
ds^2 =-2Fe^{2\psi}dv^2+2e^{\psi}dvdr+r^2d\Omega^2.\label{1}
\end{equation}
The assumption of homotheticity implies that
$F=F(x), \psi=\psi(x)$ where the similarity variable is given by $x=v/r$.

Consider the matching problem across ${\cal N}$, which we take to be the surface $V=V_0$ of ${\cal M}_1$ and the surface $v=v_0$ of ${\cal M}_2$. In analogy with the case of collapsing fluid or collisionless matter, we wish to rule out the existence of a distributional shell of matter/energy on ${\cal N}$: we impose a smooth transition on moving from ${\cal M}_1$ into ${\cal M}_2$. Spherical null shells have been discussed in \cite{BI}. The physical parameters of the shell (energy, normal forces, transverse pressure) are determined by the terms
\[ 4\pi r^2\sigma=[E],\qquad 8\pi P=[\psi_{,r}],\]
where $E=r(1-2F)/2$ is the Misner-Sharp mass of the space-time and $[\cdot]$ represents the discontinuity in a given quantity across ${\cal N}$. For ${\cal M}_1$, $E=\psi=0$ and so using the coordinates $v,r$ on ${\cal N}$, we have
\begin{eqnarray}
8\pi r\sigma&=& 1-2F|_{v=v_0}\label{2}\\
8\pi P&=&\left.-\frac{v}{r^2}\psi^\prime(x)\right|_{v=v_0}.\label{3}
\end{eqnarray}
The shell and the accompanying distributional curvature are absent iff $\sigma=P=0$. Thus from (\ref{2}) we must have
\[ F(\frac{v_0}{r})\equiv \frac12.\]
Since ${\cal N}$ extends from ${\cal J}^-$ (whereat $r=+\infty$) to the centre $r=0$, if $v_0\neq 0$, this would imply that $F(x)\equiv 1/2$ for all $x$ (without loss of generality, $v_0\geq0$). Eqn. (\ref{3}) would then give $\psi=$constant, and so ${\cal M}_2$, and hence the entire space-time,  would also be a portion of Minkowski space-time. Thus the matching must be done across $v_0=0$. This yields
\begin{equation}
F(0)=\frac12\label{4}
\end{equation}
From (\ref{3}), we see that there is {\em no} restriction on $\psi(x)$; $P$ automatically vanishes. However, we will assume henceforth that $\psi$ is constant, and without loss of generality, $\psi=0$. This is a mathematical assumption made to simplify the problem at hand, but it has a clear physical interpretation. We have
\[ 8\pi T_{ab}l^al^b=-2\frac{x}{r^2}\psi^\prime(x),\]
where ${\vec l}$ is tangent to the radial ingoing null direction. Thus by setting $\psi=0$, we are excluding the existence of such a flux: there is no pure outgoing component of field. We may refer to the line element (\ref{1}), with $\psi=0$, as the line element appropriate to an influx of field.

So we will consider the line element
\begin{equation}
ds^2=-2F(x)dv^2+2dvdr+r^2d\Omega^2,\label{5}
\end{equation}
where $x=v/r$, $v\in[0,\infty)$ and $r\in[0,\infty)$. We take $F\in C^2[0,\infty)$ in line with the usual differentiability conditions imposed on the metric tensor \cite{HE}.

\section{Energy conditions}

In spherical symmetry, the Riemann tensor and its invariants are completely specified by the following set of scalars: the Misner-Sharp energy $E$; the Newman-Penrose Weyl tensor term $\Psi_2$, calculated on a principal null tetrad; the Ricci scalar $R$ and the Ricci tensor terms $R_{\tt ii}, R_{\tt oo}$ and $R_{\tt io}$ where the subscripts ${\tt i,o}$ refer to contraction with respectively the tangent to the future directed radial ingoing and outgoing null directions. These last three are not invariant, but rescale when the radial ingoing and outgoing tangents ${\vec l},{\vec k}$ are rescaled (note however that the signs of these terms are invariant). Note that the term
\[ e^{4f}R_{\tt ii}R_{\tt oo}\]
where $e^{-2f}:=g_{ab}l^ak^b$, {\em is} an invariant, and it is this term which appears in the energy conditions.

It can be shown that the strong energy condition (SEC) (i.e. $R_{ab}u^au^b\geq 0$ for all causal $u^a$) is equivalent to
\begin{eqnarray}
R_{\tt ii}&\geq& 0,\label{ec1}\\
R_{\tt oo}&\geq&0,\label{ec2}\\
\frac12e^{2f}|R_{\tt ii}R_{\tt oo}|^{1/2}+2\frac{E}{r^3}+2\Psi_2-\frac{R}{12}&\geq&0,\label{ec3}\\
\frac12e^{2f}|R_{\tt ii}R_{\tt oo}|^{1/2}+\frac{E}{r^3}+\Psi_2-\frac{R}{6}&\geq&0.\label{ec4}
\end{eqnarray}
The weak energy condition (WEC) (i.e. $T_{ab}u^au^b\geq 0$ for all causal $u^a$) is equivalent to (\ref{ec1}), (\ref{ec2}), (\ref{ec3}) and
\begin{eqnarray}
\frac12e^{2f}|R_{\tt ii}R_{\tt oo}|^{1/2}+\frac{E}{r^3}+\Psi_2+\frac{R}{12}&\geq&0.\label{ec5}
\end{eqnarray}
The dominant energy condition (DEC) (i.e. $-T^{ab}u_b$ is future directed and causal and $T_{ab}u^au^b\geq 0$ for all future directed causal $u^a$) is equivalent to (\ref{ec1}), (\ref{ec2}), (\ref{ec3}) and
\begin{eqnarray}
\frac{E}{r^3}+\Psi_2+\frac{R}{12}&\geq&0,\label{ec6}\\
\frac12e^{2f}|R_{\tt ii}R_{\tt oo}|^{1/2}+\frac{R}{4}&\geq&0.\label{ec7}
\end{eqnarray}

For the line element (\ref{5}), we find
\begin{eqnarray*}
 E=\frac r2(1-2F),\qquad
\Psi_2=\frac{1}{6r^2}(-1+2F+4xF^\prime+x^2F^{\prime\prime}),\\
 R=\frac{2}{r^2}(1-2F+2xF^\prime-x^2F^{\prime\prime}),\qquad
R_{\tt ii}=0,\qquad R_{\tt oo}=-\frac{2}{r^2}F^\prime,
\end{eqnarray*}
where we use the scalings $l^a=\delta^a_r$ and $k^a=\delta^a_v+2F\delta^a_r$. Thus the strong energy condition is equivalent to
\begin{eqnarray}
F^\prime&\leq& 0,\label{sec1}\\
x^2F^{\prime\prime}+2xF^{\prime}-2F+1&\geq& 0,\label{sec2}\\
F^{\prime\prime}&\geq&0.\label{sec3}
\end{eqnarray}

The weak energy condition is equivalent to (\ref{sec1}), (\ref{sec2}) and
\begin{equation}
2xF^{\prime}-2F+1\geq 0.
\label{wec4}
\end{equation}

The dominant energy condition is equivalent to (\ref{sec1}), (\ref{sec2}), (\ref{wec4}) and
\begin{equation}
-x^2F^{\prime\prime}+2xF^\prime-2F+1\geq 0.\label{dec5}
\end{equation}

We will say that the function $F$ satisfies the various energy conditions if it satisfies the appropriate set of inequalities above.
The appropriate sets of conditions hold for $x\geq 0$. The consequences of the strong energy condition are quite straightforward: $F$ is non-increasing and convex. Recalling that $F(0)=1/2$, we have
\begin{equation} F(x)\leq \frac12,\qquad (x\geq 0).\label{6}\end{equation}
Integrating (\ref{sec2}) once, we find that
\[ G(x):=F^\prime +\frac{2F-1}{x}\]
is non-decreasing on $(0,\infty)$. From (\ref{sec1}) and (\ref{6}), we have $G\leq 0$ and so there exists $l\leq 0$ such that $G(x)\leq l$ and
\begin{equation}
\lim_{x\to\infty}G(x)=l.\end{equation}
Thus for any $\epsilon>0$, there exists $x_0\geq 0$ such that
\[ l-\epsilon\leq G(x)\leq l\]
for all $x\geq x_0$. Integrating from $x_0$ to $x\geq x_0$, we find that for any $\epsilon>0$, there exists $x_0\geq0$ such that
\begin{eqnarray}
&&\frac{l-\epsilon}{3}x+\frac12+\left(
F(x_0)-\frac{l-\epsilon}{3}x_0-\frac12\right)\frac{x_0^2}{x^2}
\nonumber\\
&&\leq F(x)\leq\nonumber\\
&&\frac{l}{3}x+\frac12+
\left(F(x_0)-\frac{l}{3}x_0-\frac12\right)\frac{x_0^2}{x^2},
\label{8}
\end{eqnarray}
for all $x\geq x_0$.
In particular, the limit
\[\lim_{x\to\infty} \frac{F(x)}{x}\]
exists (by which we mean exists and is finite).
We define the vector space
\[ {\cal G}=\{F\in C^2[0,\infty): \lim_{x\to\infty}\frac{F(x)}{x} {\mbox{ exists}}.\}\]
The work above shows that
\[ {\cal F}:=\{F\in C^2[0,\infty):
F(0)=1/2, F{\mbox { satisfies SEC and DEC}}\} \]
is a subset (though not a subspace) of ${\cal G}$.
In fact it is easily verified that ${\cal F}$ is a {\em convex} subset of
${\cal G}$: if $F_1,F_2\in {\cal F}$, then $\lambda F_1+(1-\lambda)F_2\in {\cal F}$ for all $\lambda\in [0,1]$.

For the remainder of this paper, we will consider the line element (\ref{5}) subject to the condition that the only non-trivial metric function $F$ is in the set ${\cal F}$. The identification of the sets ${\cal G}$ and ${\cal F}$ will allow us to put a topology on the space of solutions and consider stability issues.

Before considering the allowed global structure of the space-time, we check for the occurrence of curvature singularities. In spherical symmetry, the Kretschmann scalar is given by
\[K:=R_{abcd}R^{abcd}=24\Psi_2^2-\frac13 R^2+2R_{ab}R^{ab}.\]
For the line element (\ref{5}), this can be written in the following positive definite form:
\begin{eqnarray*} 
K&=&\frac{2}{3r^4}(1-2F-4xF^\prime-x^2F^{\prime\prime})^2+\frac{4}{3r^4}(1-2F+2xF^\prime+x^2F^{\prime\prime})^2\\
&&+\frac{4}{3r^4}(1-2F+2xF^\prime)^2+\frac{4}{3r^4}(x^2F^{\prime\prime})^2.
\end{eqnarray*}
Recalling that $x=v/r$, we note that the centre of the space-time to the future of the matching hypersurface
${\cal N}$ corresponds to $v>0, r=0$, or $v>0, x\to+\infty$. Divergences of $K$ along $v=0$ must be
ruled out, since this corresponds to the matching surface ${\cal N}$ and to (a portion of) past null
infinity. We see then that $K$ either vanishes identically, or diverges at $r=0, v>0$. The question of whether or not $K$ diverges at the boundary points corresponding to $r=v=0$ depends on the direction of approach, but typically, there will be a curvature singularity at $(v=0,r=0)$.
Lake and Zannias \cite{lake} have emphasised that these singularities are therefore purely {\em kinematical}: from the structure of $K$ above, we see that they arise from the assumption that the space-time possesses a homothetic Killing vector field. There is no dynamical input to the formation of the singularity. However, it is possible that there is a dynamical input to the {\em temporal} nature of the singularity, and it is with this issue that we will mainly concern ourselves.

\section{Classification by global structure}

The global structure of these space-times is determined by studying the null geodesic equations. The high degree of symmetry in the metric allows us to obtain first integrals for both of the non-trivial null geodesic equations; these may be written as
\begin{eqnarray*}
(F-\frac1x){\dot v}^2+\frac{v}{x^2}{\dot x}{\dot v}-\frac{x^2}{2v^2}L^2&=&0,\\
r^4{\dot x}^2+x^2(F-\frac1x)L^2&=&k,
\end{eqnarray*}
where $k$ is constant and $L$ is the conserved angular momentum. It is then reasonably straightforward to demonstrate the following results which form the basis for determining the structure of the conformal diagrams below. For brevity, we omit the proofs (the last is a straightforward application of the results of \cite{nolan99}).
This proposition determines the existence (or otherwise) and location of the apparent ($x=x_H$) and Cauchy ($x=x_0$) horizons.

\newcounter{parts}
\begin{prop}
Let $x_0$ and $x_H$ be the smallest solutions of $F(x)=1/x$ and of $F(x)=0$ respectively, if such exist. Then 
\begin{enumerate}
\item 
$x=x_H$ is the apparent horizon of the space-time; $g^{ab}\nabla_ar\nabla_br|_{x=x_H}=0$.
\item 
There are no outgoing radial null geodesics in the region $0\leq x<x_0$ which originate in the past at the singularity $(r=0,v=0)$.
\item  
$x=x_0$ is an outgoing radial null geodesic which originates in the past at the singularity $(r=0,v=0)$.
\item 
Through every point of the region $x>x_0$, there passes a 1-parameter $(L)$ family of null geodesics which originate in the past at the singularity $(r=0,v=0)$.
\item  
Every null geodesic originating at the singularity $(r=0,v=0)$ satisfies the strong singularity condition $\bullet$
\end{enumerate}
\end{prop}

Thus the qualitative behaviour of the radial null geodesics depends on the occurrence of the roots of $F$ and $F-1/x$. The energy conditions studied in Section II constrain the sets of roots to be of a few certain forms. We consider here all the possibilities which yield inequivalent conformal diagrams. Before commencing this, we take note of the following. We have $F(0)=1/2$. Since $F^\prime\leq 0$ and $F^{\prime\prime}\geq 0$ for all $x\geq 0$, we must have either $F(x)=1/2$ for $x\geq 0$, which corresponds to Minkowski space-time, or $F(x)<1/2$ for $x>0$.  We assume that the latter holds. We begin the classification by considering the occurence of an apparent horizon in the space-time. This corresponds to the existence of a solution $x_H$ of $F(x)=0$. Subject to $F^\prime\leq0$, $F^{\prime\prime}\geq 0$ there are four possibilities, which we take to define the four main classes of solution. The subclassification is based on the occurrence of naked singularities. The four classes and their respective subclasses are described now, and the corresponding conformal diagrams are given, again, for brevity, without proof.

\subsection{Class I} 
For this class, there exists $x_H>0$ such that $F(x_H)=0$ and $-\infty\leq\displaystyle{\lim_{x\to\infty}}F(x)<0$. There are two  subclasses here which we will refer to as Class Ia, where there are no NS lines ($F(x)<1/x$ for all $x\geq 0$) and Class Ib, where there exist NS lines. In the latter case, let $x_0$ and $x_1$ be respectively the smallest and largest solutions of $F=1/x$; $2<x_0\leq x_1<x_H$. We distinguish further between Class Ib1 ($x_0<x_1$) and Class Ib2 ($x_0=x_1$).

\begin{figure}
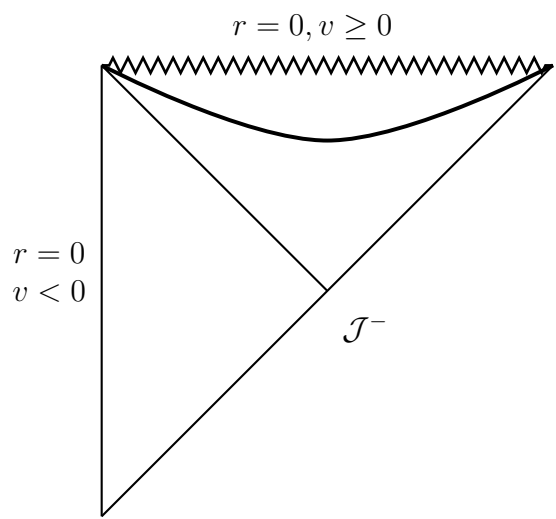

\pspicture*(-3,-3)(8,8)
$
\psline(0,0)(6,6)
\psline(3,3)(0,6)
\pscurve[linewidth=1.5pt](0,6)(3,5)(6,6)
\pszigzag[coilwidth=.2,coilarm=.1,linewidth=1pt](0,6)(6,6)
\psline(0,0)(0,6)
\rput(3.5,2.5){\psframebox*[framearc=.3]{{\cal J}^-}}
\rput(-0.7,3.5){\psframebox*[framearc=.3]{r=0}}
\rput(-0.7,3){\psframebox*[framearc=.3]{v<0}}
\rput(2.8,6.5){\psframebox*[framearc=.3]{r=0, v\geq0}}
$
\endpspicture
\caption{Conformal diagram for Class Ia. All outgoing rays cross the 
apparent horizon (bold line)
and terminate at $r=0, v\in[0,\infty)$. The singularity is censored.}
\end{figure}

\begin{figure}
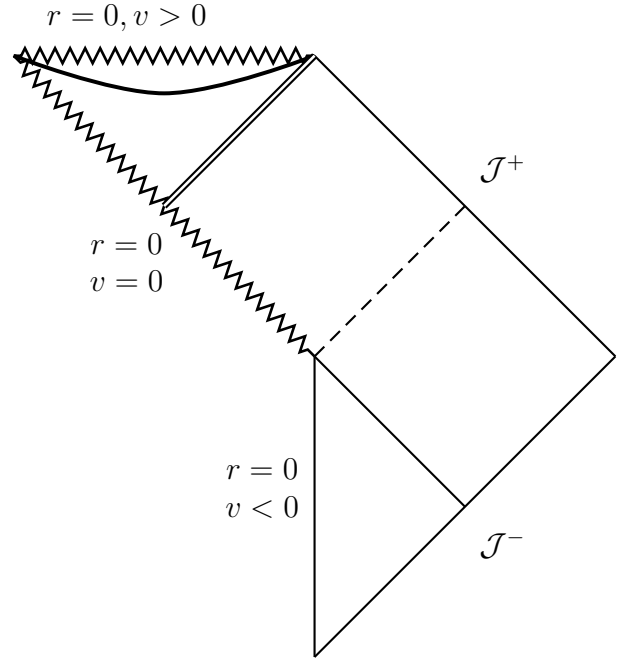

\pspicture*(-3,-3)(9,9)
$
\psline(4,0)(4,4)
\psline(4,0)(8,4)
\psline(8,4)(4,8)
\psline(6,2)(4,4)
\psline[linestyle=dashed](4,4)(6,6)
\pszigzag[coilwidth=.2,coilarm=.1,linewidth=1pt](4,4)(0,8)
\pszigzag[coilwidth=.2,coilarm=.1,linewidth=1pt](0,8)(4,8)
\pscurve[linewidth=1.5pt](0,8)(2,7.5)(4,8)
\psline[doubleline=true,doublesep=1pt](2,6)(4,8)
\rput(6.5,1.5){\psframebox*[framearc=.3]{{\cal J}^-}}
\rput(6.5,6.5){\psframebox*[framearc=.3]{{\cal J}^+}}
\rput(3.3,2.5){\psframebox*[framearc=.3]{r=0}}
\rput(3.3,2){\psframebox*[framearc=.3]{v<0}}
\rput(1.5,8.5){\psframebox*[framearc=.3]{r=0,v>0}}
\rput(1.5,5.5){\psframebox*[framearc=.3]{r=0}}
\rput(1.5,5){\psframebox*[framearc=.3]{v=0}}
$
\endpspicture
\caption{Conformal diagram for Class Ib1. The NS line $x=x_0$ (dashed) is the Cauchy
horizon of the space-time and the NS line $x=x_1$ is the event horizon 
(double line). The NS rays in the region between these two correspond to the 
globally naked portion of the singularity. The NS rays between the event horizon 
and the apparent horizon (bold) cross the latter and terminate at $r=0$. These
rays correspond to the locally naked portion of the singularity.}
\end{figure}

\begin{figure}
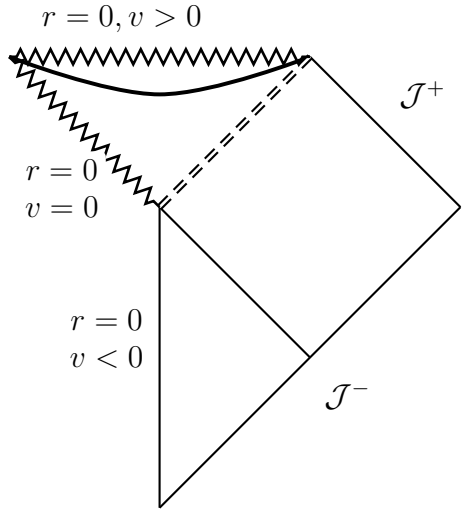

\pspicture*(-4,-4)(7,7)
$
\psline(2,0)(2,4)
\psline(2,0)(6,4)
\psline(6,4)(4,6)
\psline(4,2)(2,4)
\pscurve[linewidth=1.5pt](0,6)(2,5.5)(4,6)
\psline[linestyle=dashed,doubleline=true,doublesep=1pt](2,4)(4,6)
\pszigzag[coilwidth=.2,coilarm=.1,linewidth=1pt](2,4)(0,6)
\pszigzag[coilwidth=.2,coilarm=.1,linewidth=1pt](0,6)(4,6)
\rput(4.5,1.5){\psframebox*[framearc=.3]{{\cal J}^-}}
\rput(5.5,5.5){\psframebox*[framearc=.3]{{\cal J}^+}}
\rput(1.5,6.5){\psframebox*[framearc=.3]{r=0,v>0}}
\rput(1.3,2.5){\psframebox*[framearc=.3]{r=0}}
\rput(1.3,2){\psframebox*[framearc=.3]{v<0}}
\rput(0.7,4.5){\psframebox*[framearc=.3]{r=0}}
\rput(0.7,4){\psframebox*[framearc=.3]{v=0}}
$
\endpspicture
\caption{Conformal diagram for Class Ib2. Here, $x_0=x_1$,
so the Cauchy horizon and event horizon coincide (this is shown as a double dashed line).
The portion $(v=0,r=0)$ of the singularity is locally naked; all rays originating on the singularity
cross the apparent horizon (bold) and terminate at $r=0$.}
\end{figure}

\subsection{Class II} For Class II, there exists $x_H>0$ such that $F(x)\equiv0$ for $x\geq x_H$. The subclasses are Class IIa, where there are no roots of $F-1/x$, and Class IIb, where there exist $x_0\leq x_1<\infty$, the smallest and largest roots of $F-1/x$ respectively . 

\begin{figure}
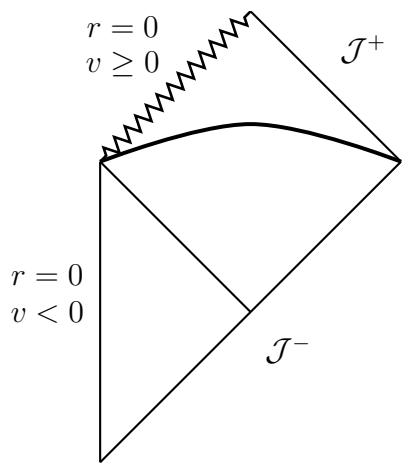

\pspicture*(-4,-4)(6,6)
$
\psline(0,0)(0,4)
\psline(0,0)(4,4)
\psline(4,4)(2,6)
\psline(2,2)(0,4)
\pscurve[linewidth=1.5pt](0,4)(2,4.5)(4,4)
\pszigzag[coilwidth=.2,coilarm=.1,linewidth=1pt](0,4)(2,6)
\rput(2.5,1.5){\psframebox*[framearc=.3]{{\cal J}^-}}
\rput(3.5,5.5){\psframebox*[framearc=.3]{{\cal J}^+}}
\rput(-0.7,2.5){\psframebox*[framearc=.3]{r=0}}
\rput(-0.7,2){\psframebox*[framearc=.3]{v<0}}
\rput(0.3,5.8){\psframebox*[framearc=.3]{r=0}}
\rput(0.3,5.3){\psframebox*[framearc=.3]{v\geq0}}
$
\endpspicture
\caption{Conformal diagram for Class IIa. All outgoing rays cross the apparent horizon
(bold) but extend out to infinity. Ingoing rays terminate along $r=0,v\geq0$.
The singularity is censored; it cannot be seen by any observer.}
\end{figure}

\begin{figure}
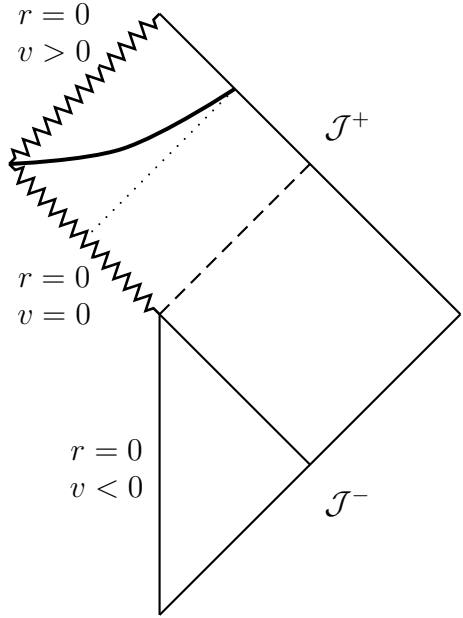

\pspicture*(-4,-4)(9,9)
$
\psline(2,0)(6,4)
\psline(2,0)(2,4)
\psline(4,2)(2,4)
\psline(6,4)(2,8)
\psline[linestyle=dashed](2,4)(4,6)
\pscurve[linewidth=1.5pt](0,6)(1.5,6.2)(3,7)
\psline[linestyle=dotted](1,5)(3,7)
\pszigzag[coilwidth=.2,coilarm=.1,linewidth=1pt](2,4)(0,6)
\pszigzag[coilwidth=.2,coilarm=.1,linewidth=1pt](0,6)(2,8)
\rput(4.5,1.5){\psframebox*[framearc=.3]{{\cal J}^-}}
\rput(4.5,6.5){\psframebox*[framearc=.3]{{\cal J}^+}}
\rput(1.3,2.2){\psframebox*[framearc=.3]{r=0}}
\rput(1.3,1.7){\psframebox*[framearc=.3]{v<0}}
\rput(0.6,8){\psframebox*[framearc=.3]{r=0}}
\rput(0.6,7.5){\psframebox*[framearc=.3]{v>0}}
\rput(0.6,4.5){\psframebox*[framearc=.3]{r=0}}
\rput(0.6,4){\psframebox*[framearc=.3]{v=0}}
$
\endpspicture
\caption{Conformal diagram for Class IIb. The NS line $x=x_0$ is the Cauchy horizon
(dashed). Outgoing rays from $(v=0,r=0)$ cross the apparent horizon (bold) iff they are 
emitted after $x=x_1$ (dotted line) which originates at $(v=0,r=0)$ and meets the apparent horizon at ${\cal J}^+$. All outgoing rays extend out to infinity. There is no 
event horizon; the singularity $(v=0,r=0)$ is globally naked and null.}
\end{figure}

\subsection{Class III} In Class III, $F(x)>0$ for all $x\geq 0$, and $\lim_{x\to\infty}F(x)=0$. There are four subclasses. For IIIa, $F(x)<1/x$ for all $x\geq 0$. For IIIb there exist $x_0,x_1$ with $x_0\leq x_1$ such that
\[ F(x)-\frac1x \left\{
\begin{array}{lll}
<0 & {\mbox {for}} & 0<x<x_0, x_1<x; \\
=0 & {\mbox {for}} & x=x_0, x=x_1. \\
\end{array}
\right.
\]
For IIIc, there exist $x_0,x_1$ with $x_0\leq x_1$ such that
\[ F(x)-\frac1x \left\{
\begin{array}{lll}
<0 & {\mbox {for}} & 0<x<x_0;\\
=0 & {\mbox {for}} & x=x_0, x=x_1;\\
>0 & {\mbox {for}} & x>x_1.\\
\end{array}
\right.
\]
For Class IIIc, the outgoing rays in $x\in[x_0,x_1]$ emanate from $(v=0,r=0)$. The rays in $x>x_1$ may originate at either $(v=0,r=0)$ or at $(v>0,r=0)$. None, all or some of these rays may originate at this latter boundary; we refer to these cases as IIIc1, IIIc2 and IIIc3 respectively. These rays are future endless and extend to infinite values of $r$.
For IIId, the set $S=\{x\in(0,\infty): F(x)=1/x\}$ is infinite and $\sup S=\infty$; there must exist $x_0=\min S \in S$.
The conformal diagrams for Class III are given in Figures 7-10.

\begin{figure}
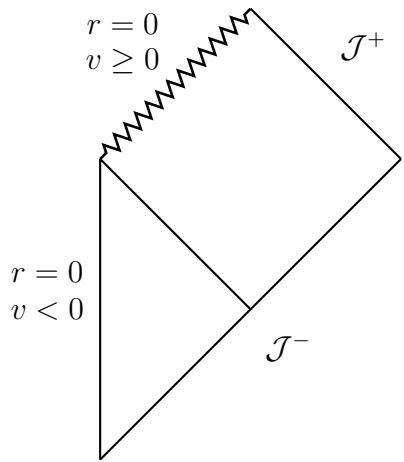

\pspicture*(-4,-4)(6,6)
$
\psline(0,0)(0,4)
\psline(0,0)(4,4)
\psline(4,4)(2,6)
\psline(2,2)(0,4)
\pszigzag[coilwidth=.2,coilarm=.1,linewidth=1pt](0,4)(2,6)
\rput(2.5,1.5){\psframebox*[framearc=.3]{{\cal J}^-}}
\rput(3.5,5.5){\psframebox*[framearc=.3]{{\cal J}^+}}
\rput(-0.7,2.5){\psframebox*[framearc=.3]{r=0}}
\rput(-0.7,2){\psframebox*[framearc=.3]{v<0}}
\rput(0.3,5.8){\psframebox*[framearc=.3]{r=0}}
\rput(0.3,5.3){\psframebox*[framearc=.3]{v\geq0}}
$
\endpspicture
\caption{Conformal diagram for Class IIIa. All ingoing rays in $v>0$ terminate on $r=0,v>0$.
There are no horizons (Cauchy, event or apparent). The singularity is censored
and is marginally trapped.}
\end{figure}

\begin{figure}
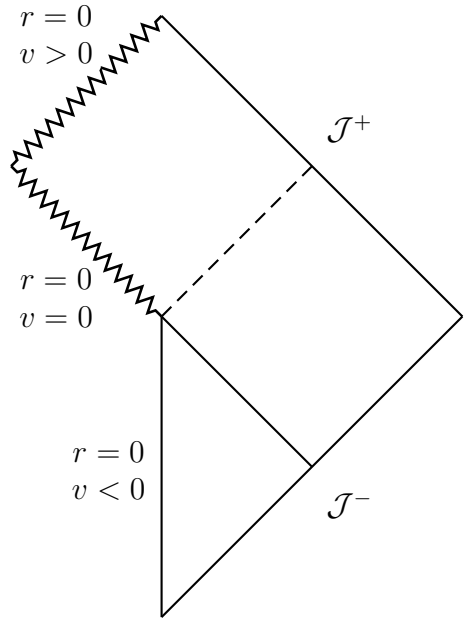

\pspicture*(-4,-4)(9,9)
$
\psline(2,0)(6,4)
\psline(2,0)(2,4)
\psline(4,2)(2,4)
\psline(6,4)(2,8)
\psline[linestyle=dashed](2,4)(4,6)
\pszigzag[coilwidth=.2,coilarm=.1,linewidth=1pt](2,4)(0,6)
\pszigzag[coilwidth=.2,coilarm=.1,linewidth=1pt](0,6)(2,8)
\rput(4.5,1.5){\psframebox*[framearc=.3]{{\cal J}^-}}
\rput(4.5,6.5){\psframebox*[framearc=.3]{{\cal J}^+}}
\rput(1.3,2.2){\psframebox*[framearc=.3]{r=0}}
\rput(1.3,1.7){\psframebox*[framearc=.3]{v<0}}
\rput(0.6,8){\psframebox*[framearc=.3]{r=0}}
\rput(0.6,7.5){\psframebox*[framearc=.3]{v>0}}
\rput(0.6,4.5){\psframebox*[framearc=.3]{r=0}}
\rput(0.6,4){\psframebox*[framearc=.3]{v=0}}
$

\endpspicture
\caption{Conformal diagram for Classes IIIb, IIIc1 and IIId. The NS line $x=x_0$ is the Cauchy horizon
(dashed). There is no apparent horizon. All outgoing rays extend out to infinity. There is no 
event horizon; the singularity $(v=0,r=0)$ is globally naked and null.}
\end{figure}

\begin{figure}
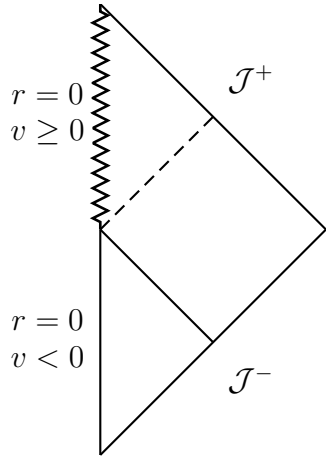

\pspicture*(-5,-5)(3,6)
$
\psline(0,0)(0,3)
\psline(0,0)(3,3)
\psline(3,3)(0,6)
\psline(1.5,1.5)(0,3)
\psline[linestyle=dashed](0,3)(1.5,4.5)
\pszigzag[coilwidth=.2,coilarm=.1,linewidth=1pt](0,3)(0,6)
\rput(2,1){\psframebox*[framearc=.3]{{\cal J}^-}}
\rput(2,5){\psframebox*[framearc=.3]{{\cal J}^+}}
\rput(-0.7,4.8){\psframebox*[framearc=.3]{r=0}}
\rput(-0.7,4.3){\psframebox*[framearc=.3]{v\geq0}}
\rput(-0.7,1.8){\psframebox*[framearc=.3]{r=0}}
\rput(-0.7,1.3){\psframebox*[framearc=.3]{v<0}}
$
\endpspicture
\caption{Conformal diagram for Classes IIIc2 and IVa. All outgoing rays extend out to infinity.
All ingoing rays in $v\geq0$ terminate on $(v\geq0,r=0)$, which is a globally
naked time-like singularity. The Cauchy horizon is dashed.}
\end{figure}

\begin{figure}
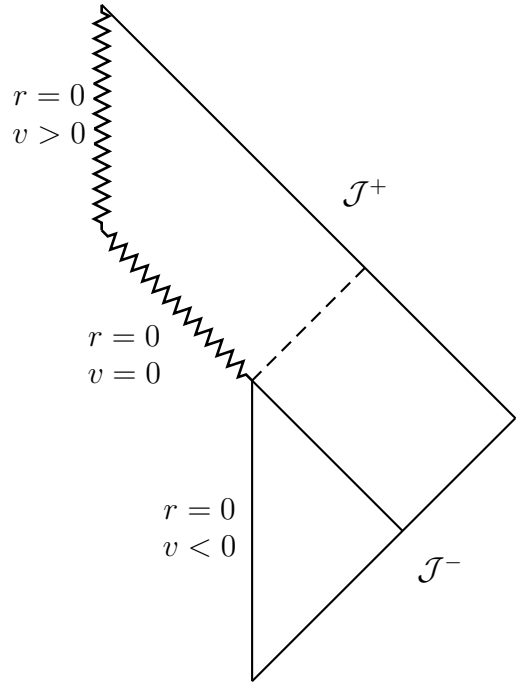

\pspicture*(-3,-3)(10,10)
$
\psline(2,0)(2,4)
\psline(2,0)(5.5,3.5)
\psline(4,2)(2,4)
\psline(5.5,3.5)(0,9)
\psline[linestyle=dashed](2,4)(3.5,5.5)
\pszigzag[coilwidth=.2,coilarm=.1,linewidth=1pt](2,4)(0,6)
\pszigzag[coilwidth=.2,coilarm=.1,linewidth=1pt](0,6)(0,9)
\rput(4.5,1.5){\psframebox*[framearc=.3]{{\cal J}^-}}
\rput(3.5,6.5){\psframebox*[framearc=.3]{{\cal J}^+}}
\rput(-0.7,7.8){\psframebox*[framearc=.3]{r=0}}
\rput(-0.7,7.3){\psframebox*[framearc=.3]{v>0}}
\rput(0.3,4.6){\psframebox*[framearc=.3]{r=0}}
\rput(0.3,4.1){\psframebox*[framearc=.3]{v=0}}
\rput(1.3,2.3){\psframebox*[framearc=.3]{r=0}}
\rput(1.3,1.8){\psframebox*[framearc=.3]{v<0}}
$
\endpspicture
\caption{Conformal diagram for Classes IIIc3 and IVb. There is no apparent horizon or event
horizon. The Cauchy horizon is dashed. The earlier portion of the singularity $(v=0,r=0)$
is null and globally naked, the later portion $(v>0,r=0)$ is time-like and globally naked.}
\end{figure}

\subsection{Class IV}
In this case, $F(x)>0$ for all $x\geq 0$, and $\lim_{x\to\infty}F(x)=l>0$.
There must exist $x_0,x_1, 0<x_0\leq x_1<\infty$ such that
\[
F(x)-\frac1x \left\{
\begin{array}{lll}
<0 & {\mbox {for}} & x<x_0;\\
=0 & {\mbox {for}} & x=x_0,x=x_1;\\
>0 & {\mbox {for}} & x>x_1.\\
\end{array}
\right.
\]
Some or all of the outoing rays originate on the singular boundary $(v=0,r>0)$. Those which do not, originate on $(v=0,r=0)$. The case where all rays originate on $(v=0,r>0)$ (call this Class IVa) is identical to Class IIIc2; the case where not all these rays originate on $(v=0,r>0)$ (call this Class IVb) is identical to Class IIIc3.
See Figures 9 and 10.

This completes the discussion of the allowed global structure of the space-times under consideration.

The most striking feature of these space-times is the common occurrence of naked singularities. Of the different classes studied above, the singularity is censored (i.e. does not admit future directed causal geodesics which terminate in the past on the singularity) for Classes Ia, IIa and IIIa only. We note that in this last case, there is no apparent horizon: this is not a necessary condition for the singularity to be censored.
Another feature which arises is that in Classes II, III and IV, the naked singularity is always {\em globally} naked. Thus there seems to be a `phase transition' between censored and globally naked singularities, which skips over locally naked singularities. However this is not the case in general, as can be seen from Class I, in the case where $x_0=x_1$. It appears that in this case the locally naked singularity solutions form a critical boundary between the censored and globally naked singularity solutions. This is borne out by the topological considerations of the following section.

In terms of significance for cosmic censorship, there are two important questions which must be addressed. Firstly, in the class of space-times being studied, are those which admit naked singularities generic, or can they be considered exceptional in some way? We address this question in the next section. Secondly, can these naked singularities arise from regular, asymptotically flat initial data? We address this question in Section 6.

\section{Stability}
In this section, we consider the topological stability of the various classes of solutions discussed above. Our main result is that the set of solutions with a globally naked singularity (GNS) may be considered to be stable in the set of all solutions under consideration: this set of solutions contains an open subset - in a natural topology - of the set ${\cal F}$. We also consider the topological properties of other sectors of the solution set. Locally naked singularity (LNS) solutions form an unstable boundary set lying between censored and GNS solutions, and while the set of censored singularity solutions is not open, the subset corresponding to a space-like singularity is open. We point out that Minkowski space-time is an interior point of the set of GNS solutions.

The space-times being studied lie in one-one correspondence with the set
\[ {\cal F}:=\{F\in C^2[0,\infty):
F(0)=1/2, F{\mbox { satisfies SEC and DEC}}\}, \]
which as pointed out above, is a convex subset of the real vector space
\[ {\cal G}=\{F\in C^2[0,\infty): \lim_{x\to\infty}\frac{F(x)}{x} {\mbox{exists}}.\}\]
The defining property of ${\cal G}$ suggests a suitable norm (from which a topology is inherited in the usual way). We define $\|\cdot\|:{\cal G}\to{\bf R}^+$ by
\[ \|F\|=\sup_{x\in[0,2)}|F|+\sup_{x\in[2,\infty)}|\frac{F}{x}|.\]
It is easily verified that this defines a norm on ${\cal G}$. The cut-off at $x=2$ could be done at any $x=\epsilon>0$, but is made at $x=2$ for convenience: intersections of $y=1/x$ and $y=F(x)$, which correspond to naked singularity lines can only arise if $x\in[2,\infty)$ and it is convenient to have to deal with just one of the elements of the formula for the norm at such points.

We define ${\cal F}_{NS}$ and ${\cal F}_{CEN}$ to be the subsets of ${\cal F}$ which correspond respectively to solutions which admit  naked singularities and those which do not. From the previous section, we can write
\[ {\cal F}_{CEN}=\{ F\in {\cal F}: F(x)<\frac1x {\mbox { for all } }x>0\}.\]
${\cal F}_{CEN}$ and ${\cal F}_{NS}$ form a partition of ${\cal F}$.
We introduce the following decomposition: ${\cal F}_{NS}={\cal F}_{NS1}\sqcup {\cal F}_{NS2}$ (disjoint union) where
\begin{eqnarray*}
{\cal F}_{NS1}=\{&&\!\!\!F\in {\cal F}_{NS}: {\mbox {there exist } }x_0\in(0,\infty), x_1>x_0 {\mbox { s.t. }}\\ && F(x_0)=\frac{1}{x_0}, F(x_1)>\frac{1}{x_1}\},
\end{eqnarray*}
\[ {\cal F}_{NS2}=\{F \in {\cal F}_{NS}: {\mbox {there exists } }S\subset {\bf R} {\mbox { s.t. } }
F(x)\left\{
\begin{array}{ll}
=1/x,& x\in S;\\
<1/x, & x\not\in S.\\
\end{array}
\right.
\}
.\]
Notice that all $F\in{\cal F}_{NS1}$ correspond to GNS space-times, while all LNS spacetimes correspond to $F\in{\cal F}_{NS2}$. The important topological features, in terms of the existence of naked singularities, of the set of solutions ${\cal F}$, are given by the following list of results.

\begin{prop}
${\cal F}_{NS1}$ is an open subset of ${\cal F}$.
\end{prop}

{\bf Proof:} Let $F\in{\cal F}_{NS1}$ and let $x_0,x_1$ be as in the definition of ${\cal F}_{NS1}$. Let
\[ 0<\epsilon<\frac{1}{x_1}(F(x_1)-\frac{1}{x_1}).\]
Consider the neighbourhood of $F$ in ${\cal F}$ defined by $\|F-G\|<\epsilon$. For any $G$ in this neighbourhood, we have
\[ \sup_{x\in[0,2)}|F-G| + \sup_{x\in[2,\infty)}|\frac{F-G}{x}|<\epsilon.\]
Thus for $x\geq 2$,
\[ F-\epsilon x<G<F+\epsilon x.\]
In particular,
\begin{eqnarray*}
G(x_1)&>&F(x_1)-\epsilon x_1\\
&>&\frac{1}{x_1}.
\end{eqnarray*}
Thus $G\in {\cal F}_{NS1}$ and so ${\cal F}_{NS1}$ is open$\bullet$

We denote the topological boundary of a set $A$ by $b(A)$.

\begin{prop}
${\cal F}_{NS2}\subseteq b({\cal F}_{NS})$.
\end{prop}

{\bf Proof:} Let $F\in {\cal F}_{NS2}$ and let $x_0$ be the least element of $S$ ($S$ is a root set of a continuous function and so $x_0\in S$ exists). Let $\epsilon>0$ and let $G_-=F-\epsilon^\prime x/3$, where $0<\epsilon^\prime<\epsilon$. Then it is easily verified that $G_-\in {\cal F}$. Furthermore, it is clear that $G_-\in {\cal F}_{CEN}$. Also
\begin{eqnarray*}
\|F-G\|&=&\|\frac{\epsilon^\prime x}{3}\|\\
&=&\sup_{x\in[0,2)}|\frac{\epsilon^\prime x}{3}|+\sup_{x\in[2,\infty)}|\frac{\epsilon^\prime }{3}|\\
&=&\epsilon^\prime<\epsilon.
\end{eqnarray*}

For $k>0$, define
\[ G_+ =kF+\frac12(1-k).\]
By convexity, $G_+\in{\cal F}$ for $k\in(0,1]$ (recall that $F=1/2\in{\cal F}$ corresponds to Minkowski space-time).
Now $G_+(x_0)>1/x_0$ iff $k<1$ (we know that $x_0>2$). So for $k<1$, $G_+\in {\cal F}_{NS1}$. Also
\begin{eqnarray*}
\|G-F\|&=&\|(k-1)(F-\frac{1}{2})\|\\
&=&(1-k)\|F-\frac12\|,
\end{eqnarray*}
which is strictly bounded above by $\epsilon$ provided $k$ is sufficiently close to 1 ($F$ is fixed and so $\|F-\frac12\|$ is a fixed number).

Thus for any $\epsilon>0$, the $\epsilon-$neighbourhood of $F$ in ${\cal F}$
intersects both ${\cal F}_{CEN}$, which by definition, is the
${\cal F}-$complement of ${\cal F}_{NS}$, and ${\cal F}_{NS1}$,
which being open, is a subset of the interior of ${\cal F}_{NS}$.
Thus $F\in b({\cal F}_{NS})$. This completes the proof$\bullet$

We introduce the following partition of the subset of ${\cal F}$ corresponding to censored singularities:${\cal F}_{CEN}={\cal F}_{CEN1}\sqcup {\cal F}_{CEN2}$, where
\begin{eqnarray*} {\cal F}_{CEN1}=\{&&\!\!\!F\in{\cal F}_{CEN}: {\mbox {there exists }} x_H>0 {\mbox { s.t. }}\\
&& F(x_H)=0, F(x)<0 {\mbox { for }} x>x_H\},
\end{eqnarray*}
\[
{\cal F}_{CEN2}={\cal F}_{CEN}-{\cal F}_{CEN1}.
\]

\begin{prop}
${\cal F}_{CEN1}$ is an open subset of ${\cal F}$.
\end{prop}

{\bf Proof:} Let $F\in{\cal F}_{CEN1}$, let $x_H$ be as in the definition and let $x_1>x_H$. Define $l_1=|F(x_1)|>0$ and let $\epsilon_1=l_1/x_1>0$. If $G\in{\cal F}$ and $\|F-G\|<\epsilon_1$, then
\[ |\frac{F-G}{x}|<\epsilon_1, {\mbox { for all }} x\geq 2,\]
so that
\[ G(x)<F(x)+\epsilon_1 x,\qquad x\geq 2.\]
In particular, $G(x_1)<F(x_1)+\epsilon_1 x_1=0$, so that $G$ has a root; call this root $x_2$.
Let
\[ l_2=\min_{x\in[2,x_2]} |F-\frac1x|.\]
By definition of ${\cal F}_{CEN}$, $l_2$ is strictly positive. Thus
\[ F(x)\leq \frac1x - l_2,  {\mbox { for all }} x\in[2,x_2].\]
Let $0<\epsilon_2<l_2/x_2$. For $G\in{\cal F}$ with $\|F-G\|<\epsilon_2$, we have that for $x\in[2,x_2]$
\[ |\frac{F-G}{x}|<\epsilon_2,\]
so that
\begin{eqnarray*}
G&<&F+\epsilon_2 x\\
&\leq&\frac1x+\epsilon_2 x -l_2\\
&<&\frac1x,
\end{eqnarray*}
where the last line comes about by the definition of $\epsilon_2$.
Thus for $\epsilon=\min\{\epsilon_1,\epsilon_2\}$, $\|F-G\|<\epsilon$ guarantees that $G\in{\cal F}_{CEN}$, proving that ${\cal F}_{CEN}$ is an open subset of ${\cal F}$ $\bullet$

\begin{prop}
${\cal F}_{CEN2}\subseteq b({\cal F}_{CEN})$.
\end{prop}

{\bf Proof:}
Let $F\in {\cal F}_{CEN2}$ and let $\epsilon>0$. Let $0<\epsilon^\prime<\epsilon$ and define
\[ G_-=F-\frac{\epsilon^\prime}{3}x.\]
Then it is easily verified that $G_-\in {\cal F}$ and by definition of ${\cal F}_{CEN2}$,
\[G_-<\frac1x-\frac{\epsilon^\prime}{3}x<\frac{1}{x}\]
for all $x>0$. It is clear that there exists $x_H>0$ such that $G_-(x_H)=0$ and $G_-(x)<0$ for $x>x_H$. Thus $G_-\in{\cal F}_{CEN1}$.
Furthermore, $\|F-G_-\|=\epsilon^\prime<\epsilon$, so $G_-$ lies in the $\epsilon-$neighbourhood of $F$.

Define $G_+=kF+(1-k)/2$. By convexity, $G_+\in{\cal F}$ for $k\in[0,1]$. Since $F\in{\cal F}_{CEN2}$, we have
\[ \lim_{x\to\infty} G_+(x)=\frac{1-k}{2},\]
which is positive provided $k<1$. So for $k\in(0,1), G_+\in{\cal F}_{NS}$. Also, $\|F-G_+\|=(1-k)\|F-1/2\|$, which for any $F\in{\cal F}_{CEN2}$ and $\epsilon>0$ may be made less than $\epsilon$ by taking $k$ to be sufficiently close to 1.

Thus every $\epsilon-$neighbourhood of $F\in{\cal F}_{CEN2}$ intersects both the interior and the complement of ${\cal F}_{CEN}$, proving the assertion$\bullet$

This gives a complete decomposition of the solution space into regions where the associated space-times admit naked and censored singularities. Both these sectors are stable in the sense that there are open subsets of both NS solutions and censored singularity solutions. Two other points of interest arise.

First, we see that the element of ${\cal F}$ corresponding to Minkowski space-time, namely $F_M:=F\equiv 1/2$, is an interior point of ${\cal F}_{NS}; F_M\in{\cal F}_{NS1}$. So in this class of solutions, there are arbitrarily small perturbations of Minkowski space-time which admit naked singularities. This feature is borne out when we consider these solutions from the point of view of the initial value problem.

Secondly, we see that those solutions which admit locally, but not globally naked singularities, are unstable. These solutions correspond to the subset of ${\cal F}_{NS2}$ for which elements $F$ also have an isolated root, i.e. for which there exists $x_H$ such that $F(x_H)=0$ and $F(x)<0$ for $x>x_H$. This describes the solutions of Class Ib for which $x_0=x_1$. Being a subset of the boundary set ${\cal F}_{NS2}$, these solutions may be considered to form an unstable critical boundary between globally naked solutions and censored solutions.

These results describe a hierarchy of space-times, starting with Minkowski space-time and moving through globally naked and locally naked singularity space-times to space-times admitting only censored singularities. This scenario is in violation of some formulations of the cosmic censorship hypothesis.

\section{The initial value perspective}
The cosmic censorship hypothesis (CCH) addresses the nature of singularities which arise in the evolution of space-time, as governed by Einstein's equation, from regular, asymptotically flat initial data \cite{waldrev}. Thus demonstrating the existence of solutions of Einstein's equation without reference to initial data does not necessarily offer any evidence against cosmic censorship. In order to determine if the solutions studied here have any significance for the CCH, we consider how they arise as the development of initial data. We claim that the solutions studied here can indeed arise from regular, asymptotically flat intial data, albeit data of rather low differentiability. As the space-times are self-similar, a non-singular space-like slice is not naturally asymptotically flat. Asymptotic flatness is imposed by matching the self-similar region of an initial data slice with a slice of the exterior Schwarzschild solution. The full matching problem, i.e. that of joining an interior self-similar region with a vacuum exterior across a time-like hypersurface will not be considered here. It is sufficient for our purposes to consider the matching of the initial data surface.

In the space-time region prior to the Cauchy horizon (CH), the similarity variable $x=v/r$ provides a natural time coordinate; we have
\[ g^{ab}\nabla_ax\nabla_bx=2\frac{x^2}{r^2}(F-\frac1x).\]
If the CH occurs at $x=x_{CH} (x_{CH}=+\infty$ for censored singularities), then the slice $\Sigma_i:=\{x:x=x_i<x_{CH}\}$ is a space-like hypersurface with unit future-pointing time-like normal
\[ n^a=\frac{1}{\sqrt{2}x_i}(\frac{1}{x_i}-F_i)^{-1/2}\{x_i\delta^a_v-(1-2x_iF_i)\delta^a_r\}.\]
However such a surface {\em does not} provide regular initial data surface, as it includes the point $(v=0,r=0)$ and consequently scalar invariants of the Riemann tensor will be infinite on $\Sigma_i$. This problem can be avoided by taking the initial data surface to be
\[ \Sigma^\prime_i=\{x^a\in {\cal M}:v=\left\{
\begin{array}{ll}
x_ir,&\epsilon\leq r<\infty;\\
\lambda(r),&0\leq r<\epsilon.
\end{array}
\right.
\},
\]
where $\lambda, \epsilon$ are chosen so that $\Sigma^\prime_i$ is $C^\infty$, space-like and intersects $r=0$ at $v<0$, i.e. in the flat region of space-time. See Figure 11.

\begin{figure}
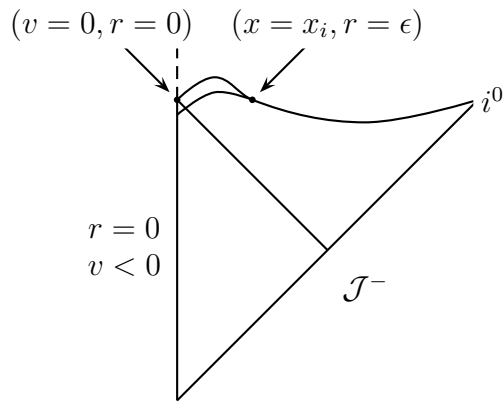

\pspicture*(-4,-4)(7,7)
$
\psline(0,0)(0,4)
\psline(0,0)(4,4)
\psline(2,2)(0,4)
\psline[linestyle=dashed](0,4)(0,5)
\pscurve(0,4)(0.5,4.3)(1,4)(2.5,3.7)(4,4)
\pscurve(0,3.8)(0.5,4.1)(1,4)
\psline[arrows=<-](1.1,4.1)(1.8,4.8)
\rput(2,5){\psframebox*[framearc=.3]{(x=x_i,r=\epsilon)}}
\rput(4.2,4){\psframebox*[framearc=.3]{i^0}}
\rput(2.5,1.5){\psframebox*[framearc=.3]{{\cal J}^-}}
\rput(-0.7,2.3){\psframebox*[framearc=.3]{r=0}}
\rput(-0.7,1.8){\psframebox*[framearc=.3]{v<0}}
\psdots*(1,4)
\psdots*(0,4)
\psline[arrows=<-](-0.1,4.1)(-0.8,4.8)
\rput(-1,5){\psframebox*[framearc=.3]{(v=0,r=0)}}
$
\endpspicture
\caption{Diagram describing the initial data surfaces $\Sigma_i$ and $\Sigma^\prime_i$.
$\Sigma_i$ joins $(r=0,v=0)$ and $i^0$ and is given by $x=x_i$. $\Sigma^\prime_i$
agrees with $\Sigma_i$ for $r\geq\epsilon$; for $r<\epsilon$, $\Sigma^\prime_i$
is given by $v=\lambda(r)$ and lies to the past of the corresponding portion of $\Sigma_i$. 
$\Sigma^\prime_i$ intersects the centre $(r=0)$ at a regular point of space-time
at which $v<0$.}
\end{figure}

As seen in Section 2, the space-times being studied inevitably have a curvature singularity at $(v=0,r=0)$. This is a consequence of the assumption of homothetic symmetry (self-similarity), and may be viewed as being purely kinematical. In particular, it {\em cannot} be ascribed to the magnitude of the initial data on $\Sigma^\prime_i$; these data (as described below) inevitably lead to a curvature singularity (except in the case of trivial data).  Since these data {\em implicitly} contain a singularity, they may be considered to be on the same footing (as regards the initial value problem) as the more easily accessible data prescribed on $\Sigma_i$, which are {\em explicitly} singular. The specification of the regular data on $\Sigma^\prime_i$ and the assumption of self-similarity completely and uniquely determine the singular data on $\Sigma_i$. From this point of view, we may consider the data on $\Sigma_i$ to be regular.
Thus, apart from the question of asymptotic flatness, which we will deal with later, these data are appropriate for the consideration of the question of cosmic censorship, i.e. the question of how the {\em temporal} nature of the singularity relates to the initial data.

The initial data on $\Sigma_i$ consist of the 3-metric $h_{\alpha\beta}$, the extrinsic curvature $K_{\alpha\beta}$ and the initial data for the matter fields which generate the geometry. As we have not specified the matter distribution of the space-time, we cannot specify the last of these. However these data are not freely specifiable, as they are subject to the initial value constraints,
\[ G_{ab}n^b=8\pi T_{ab}n^b,\]
and so the free initial data consist of $h_{\alpha\beta}$ and $K_{\alpha\beta}$. In local coordinates $\xi^\alpha=(r,\theta,\phi)$ on $\Sigma_i$, these have components
\begin{eqnarray}
h_{\alpha\beta}&=&{\mbox {diag}}(2x_i(1-x_iF_i),r^2,r^2\sin^2\theta),\label{3met}\\
K_{\alpha\beta}&=&{\mbox {diag}}(2\frac{x_i^3}{\kappa r}F^\prime_i(x_iF_i-1),\frac{r}{\kappa}
(2x_iF_i-1),\frac{r}{\kappa}(2x_iF_i-1)\sin^2\theta),
\label{3ext}
\end{eqnarray}
where $F_i=F(x_i), F^\prime_i=F^\prime(x_i)$ and $\kappa^2=2x_i(1-x_iF_i)$ ($\kappa^2>0$ since $x_i<x_{CH}$.)
We note that the Ricci scalar of $\Sigma_i$ is given by
\[ ^{(3)}\!R=-\frac{(1-2x_iF_i)(1-2x_iF_i+2x^2F^\prime_i)}{r^2x_i(1-x_iF_i)}.\]

The trivial data, i.e. those corresponding to $\Sigma_i$ being a slice of Minkowski space-time, have $F_i=1/2, F^\prime_i=0$. With these data, the energy conditions (which necessitate $F^\prime\leq0$, $F^{\prime\prime}\geq 0$) yield $F\equiv1/2$, and so the entire space-time is flat. Notice however that even for these data, $\Sigma_i$  is singular both at the origin $r=0$, and in the limit as $\Sigma_i$ approaches a null hypersurface (i.e. as $x_i$ approaches the first root of $F-1/x$). This occurs at $x=2$ for the trivial case. This is an indication that these features of general initial data are unimportant.

Our aim now is to write down an appropriate measure for the initial data. Subject to the assumptions made thus far, we see that the data on a fixed $\Sigma_i$ are completely specified by the two parameters $F_i$ and $F^\prime_i$. These are subject to the constraints $F_i\leq 1/2$, $F^\prime_i\leq 0$ and the first order energy condition $(\ref{wec4})$. These are the only restrictions on the data $\{F_i,F^\prime_i\}$, and we will assume henceforth that these inequalities are satisfied. The trivial data are $F_i=1/2, F^\prime_i=0$. Thus instead of using a standard measure of initial data as in \cite{HE} or \cite{klain} (which in any case would not be well defined for these singular hypersurfaces), we will use
\[ \mu_i = \max\{\frac12-F_i,-F^\prime_i\}.\]
Note that $\mu_i\geq0$ and that $\mu_i=0$ iff the data are trivial.

The following results describe connections between the magnitude of the initial data described here and the temporal nature of the singularity. The first result gives a sufficient condition on the initial data for the singularity to be naked.

\begin{prop}
For every $x_i>0$ and $0<F_i<1/x_i$, there exists $\epsilon_i>0$ such that if the initial data $\{F_i,F^\prime_i\}$ satisfy $\mu_i<\epsilon_i$, then the corresponding space-time contains a naked singularity.
\end{prop}

{\bf Proof:} A sufficient condition for the graph of $F$, passing through $(x_i,F_i)$, to intersect the graph of $1/x$ is that for some $x_1>x_i$, $F$ lies above the line through $(x_i,F_i)$ which is tangent to $1/x$ at $x_1$. So let $x_1>2$ be arbitrary. Equating the slope of the line joining $(x_i,F_i)$ and $(x_1,1/x_1)$ with the tangent slope $-1/x_1^2$, and taking the appropriate root of the resulting quadratic equation gives
\[ x_1=\frac{1+(1-x_iF_i)^{1/2}}{F_i}.\]
Then since $F^\prime\leq 0$ and $F^{\prime\prime}\geq0$, a sufficient condition for $F$ to intersect $1/x$ is that
\begin{equation} |F^\prime_i|<\frac{1}{x_1^2}=\frac{F_i^2}{(1+(1-x_iF_i)^{1/2})^2}.
\label{ivp1}
\end{equation}
We show that for each $x_i$, by taking $\epsilon_i>0$ to be sufficiently small, this condition is satisfied.

So let $\epsilon>0$ and assume that $\mu_i<\epsilon$. Then $F_i>-\epsilon+1/2$ and $|F^\prime_i|<\epsilon$. The inequality (\ref{ivp1}) is satisfied if
\[ \delta(1+(1-x_iF_i)^{1/2})<F_i,\]
where $\delta=\sqrt{\epsilon}$, so (\ref{ivp1}) holds if
\begin{equation}
(1-x_iF_i)^{1/2}<\frac{F}{\delta}-1.\label{ivp2}
\end{equation}
The condition $F_i>-\epsilon+1/2$ guarantees that the right hand side of (\ref{ivp2}) is positive provided
\begin{equation} \delta<\frac{\sqrt{3}-1}{2},\label{ivp3}\
\end{equation}
which restriction we assume henceforth.

Then (\ref{ivp2}) holds iff
\begin{equation}
F_i>2\delta-\delta^2x_i.\label{ivp4}
\end{equation}
Again $F_i>-\epsilon+1/2$ will imply this inequality, provided that
\begin{equation}
(1-x_i)\delta^2+2\delta-\frac12<0.\label{ivp5}
\end{equation}
The left hand side here is a continuous function of $\delta$ which
is negative at $\delta=0$. Thus for all $x_i>0$, there is
an interval's worth $(0,\delta_{i0})$ of positive values of $\delta$
for which (\ref{ivp5}) is valid. Taking $\epsilon_i$ to be the smaller
of $\delta_{i0}^2$ and $(\frac{\sqrt{3}-1}{2})^2$ proves the conclusion
of the theorem$\bullet$

The following result provides a sufficient condition for the singularity to be censored.

\begin{prop}
Given $x_i<1$, if the initial data $\{F_i>0,F^\prime_i\}$ satisfy  
\[ |F^\prime_i|>\max\{\frac{1}{4x_i},\frac{F_i}{x_i}\}\]
then the singularity is censored.
\end{prop}

{\bf Proof:}
The proof relies heavily on the energy conditions. Remember that a sufficient condition for the singularity to be censored is that there exists $x_0>0$ such that $F<1/x$ on $[x_i,x_0)$ and $F(x_0)<0$. Integrating the dominant energy condition (\ref{dec5}), we see that $g(x)$ (defined below) is non-increasing; (\ref{sec1}) and (\ref{wec4}) show $g$ to be non-negative. Thus
\[ 0\leq g(x):= F^\prime(x)+\frac{1-2F(x)}{x}\leq F^\prime_i+\frac{1-2F_i}{x_i}\]
for all $x\geq x_i$.
Integrating this inequality over the interval $[x_i,x]$, we obtain
\[ F(x)\leq h(x):=\left(F^\prime_i+\frac{1-2F_i}{2x_i}\right)\frac{x^2}{x_i}-\left(F^\prime_i+\frac{1-2F_i}{x_i}\right)x+\frac12.\]
To complete the proof, we show that there exists $x_0>0$ such that $h<1/x$ for $x\in[x_i,x_0]$ and $h(x_0)<0$. Define $u(x)=xh(x)-1$. Then $h(x)<1/x$ iff $u(x)<0$. Note that $u$ is cubic in $x$ with positive coefficent of $x^3$. We calculate
\begin{eqnarray*}
u(x_i)&=&x_iF_i-1<0,\\
u^\prime(x_i)&=&x_iF^\prime_i+F_i<0,\\
u(2x_i)&=&4x_i^2F^\prime_i+x_i-1<0,
\end{eqnarray*}
where the second and third inequalities come about from the hypotheses of the Proposition. Since $u$ is a cubic function, these conditions are sufficient to guarantee that $u<0$ on $[x_i,2x_i]$.
To complete the proof, we note that
\[ h(2x_i)=2x_iF^\prime_i+\frac12,\]
which is negative by hypothesis$\bullet$

Notice that the initial data here include the `no trapped surfaces' condition $F_i>0$. We note also that the hypotheses constrain $F_i$ to satisfy $F_i<1/3$, and so we see that the result says that if the initial data are sufficiently large (as measured by $\mu_i$), then the singularity will be censored.

We have not investigated the nature of initial data lying in the region $\epsilon_i\leq \mu_i\leq \lambda_i$. In particular, we have not checked for the presence of critical boundaries in the initial data space which may mark the transition from globally naked to locally naked singularities, or from naked to censored singularities.
In the absence of a full theory, i.e. a specification of the matter distribution of the space-time, it does not make sense to do so, as it is possible that the same initial data $(F_i,F^\prime_i)$ would evolve differently (naked or censored singularity) depending on the matter distribution. For example, if we specify that the energy-momentum tensor is that of a null fluid (in which case the space-time is the self-similar Vaidya space-time), then we have
\[ F(x)=\frac12-\gamma x,\quad \gamma>0.\]
The singularity is censored if $\gamma>1/16$, locally naked if $\gamma=1/16$ and globally naked if $\gamma<1/16$. In the critical case, the straight line graph of $F$ is tangent to $1/x$. Thus the critical value of $\gamma$ is $1/16$, which translates into critical values of the initial data $F_i=1/2-\gamma x_i, F^\prime_i=-\gamma$. Now consider these data for any other matter model in which $F^{\prime\prime}\neq 0$. The non-vanishing second derivative will cause the graph of $F$ to bend upwards ($F^{\prime\prime>}0$) from the straight line Vaidya position, and so the initial data which correspond to the critical Vaidya solution will lead to a globally naked singularity in such a matter model.

The propositions of this section reinforce the hierarchical interpretation set out in the previous section. Again, the solutions close to Minkowski space-time - this time in the sense of the data being arbitrarily close to trivial - admit globally naked singularities, while the singularities in those space-times which are sufficiently far from Minkowski space-time are censored.

One issue remains; the data sets studied here are not asymptotically flat. In the remainder of this section, we show that the intial data surface $\Sigma_i$ may be matched at a finite radius to an asymptotically flat initial data surface. The resulting initial data sets {\em are} asymptotically flat, and reproduce the behaviour described by Propositions 6 and 7 above.

So we consider this situation. The new initial data surface is formed by taking a space-like slice through the space-time formed by matching a self-similar interior region as described by (\ref{5}) with $F\in{\cal F}$ across a time-like boundary with the exterior Schwarschild solution. Taking the boundary of the infalling homothetic field and the vacuum region to intersect this  slice $\Sigma_{if}$ at $r=r_0$, the slice will thus consist of an interior portion of $\Sigma_i$, joined across $r=r_0$ with a space-like slice of the exterior Schwarzschild solution. We write the line element for the exterior Schwarzschild field using the advanced time null coordinate $w$;
\[ds^2 = -\bigtriangleup dw^2 +2dwdr+r^2d\Omega^2,\]
where $\bigtriangleup=1-2m/r$ and $m$ is the Schwarzschild mass parameter. We consider the hypersurface $\Sigma_{ext}$ given by $w=h(r)$, where $h\in C^2[r_0,\infty)$. This hypersurface is space-like iff $\rho^2:=2h^\prime-\bigtriangleup(h^\prime)^2>0$, which is assumed henceforth. In coordinates $\xi^\alpha=(r,\theta,\phi)$, the components of the 3-metric and extrinsic curvature of $w=h(r)$ are respectively
\begin{eqnarray}
h^+_{\alpha\beta}&=&{\mbox{diag}}(\rho^2,r^2,r^2\sin^2\theta),\label{smet}\\
K^+_{\alpha\beta}&=&{\mbox{diag}}(\frac{h^{\prime\prime}}{\rho}+\frac{m(h^\prime)^2}{\rho r^2}(3-\bigtriangleup h^\prime),\frac{r}{\rho}
(\bigtriangleup h^\prime-1),\frac{r}{\rho}
(\bigtriangleup h^\prime-1) \sin^2\theta).\label{sext}
\end{eqnarray}

Thus the initial data surface $\Sigma_{if}$ consists of the union  of $\Sigma_i$ for $0\leq r\le r_0$ and $\Sigma_{ext}$ for $r_0\leq r\le\infty$. The initial data are the 3-metric
${\bar h}_{\alpha\beta}$ and extrinsic curvature ${\bar K}_{\alpha\beta}$ , whose components in the coordinates $\xi^\alpha=(r,\theta,\phi)$ are
\begin{eqnarray*}
{\bar h}_{\alpha\beta}&=&\left\{
\begin{array}{ll}
h_{\alpha\beta},&0\leq r<r_0;\\
h^+_{\alpha\beta}, &r_0\leq r<\infty,\\
\end{array}
\right.
\\
{\bar K}_{\alpha\beta}&=&\left\{
\begin{array}{ll}
K_{\alpha\beta},&0\leq r<r_0;\\
K^+_{\alpha\beta}, &r_0\leq r<\infty.\\
\end{array}
\right.
\end{eqnarray*}

Demanding various degrees of smoothness of ${\bar h}_{\alpha\beta}$ and ${\bar K}_{\alpha\beta}$ across $r=r_0$ will put constraints on $x_i, F_i, F^\prime_i$. We consider the effect of these constraints for the existence or otherwise of the various classes of solutions found above, in particular for those solutions admitting naked singularities. Some straightforward algebra yields the following.

\begin{prop} Consider the initial data surface $\Sigma_{if}$
consisting of an interior portion lying in the self-similar region
with line element (\ref{5}) with $F\in{\cal F}$, truncated at the
origin to avoid the singularity, and an exterior portion lying in
the Schwarzschild space-time. Then $\Sigma_{if}$ cannot have both
a 3-metric which is differentiable and an extrinsic curvature
tensor which is continuous for all $r>0$ $\bullet$
\end{prop}

Thus demanding this degree of regularity for the initial data
constructed above rules out {\em all} of the solutions being
considered here, in particular those containing naked
singularities. This is a rather harsh version of cosmic censorship
for these space-times.

We now consider the situation where the 3-metric and extrinsic
curvature are only assumed to be continuous.  For convenience, we introduce the constant $\alpha$ and write
\[h^{\prime\prime}(r_0)=\alpha\frac{m}{r_0^2}x_i^2.\]
The range of values of $\alpha$ allowed by the energy conditions on the initial data surface is
 \begin{equation} 2x_iF_i < \alpha+3\leq
2.\label{m10}
\end{equation}
(The strict inequality arises by
imposing strict inequality in (\ref{sec1}) which has the effect of
ruling out the trivial situation $F^\prime_i=0$.) Within this range of allowed values, both naked
singularity solutions and censored singularity solutions survive.
The following propositions are direct applications of Propositions 6 and 7 respectively.

\begin{prop}
For every $0<x_i<2$, if $m/r_0>0$ is sufficiently small, then all
space-times which evolve from the initial data set
$\{\Sigma_{if},{\bar h}_{\alpha\beta},{\bar K}_{\alpha\beta}\}$
subject to continuity, but not differentiability of ${\bar
h}_{\alpha\beta}$ and ${\bar K}_{\alpha\beta}$, admit naked
singularities $\bullet$

\end{prop}

\begin{prop}
Let $x_i<1, 0<F_i<1/4$. Then there is an interval of allowed values of $\alpha$ such that the continuous initial data $\{{\bar h}_{\alpha\beta},{\bar K}_{\alpha\beta}\}$ evolve to a censored singularity $\bullet$
\end{prop}

\section{Conclusions}
We have classified, in terms of their global structure, all spherically symmetric self-similar space-times of a particular form. The physical significance of these rests on the fact that they satisfy the strong and dominant energy conditions. We have shown that the set of solutions admitting naked singularities and that admitting censored singularities are both stable in the sense of containing open subsets in the space of solutions. We have also identified a portion of the critical boundary separating these sets; this portion corresponds to space-times admitting locally naked singularities.
Furthermore, we have given examples of space-times which arise from regular asymptotically flat initial data and which nonetheless admit both globally and locally naked singularities. Those solutions which admit naked singularities arise when the data are sufficiently small. Thus naked singularities can arise from initial data which come about due to a small perturbation of the trivial data, for which the evolution leads to Minkowski space-time. The degree of `regularity' of the initial data is quite low; if the 3-metric is assumed to be differentiable throughout the chosen slice, then the energy conditions are violated and the solutions are ruled out. 

One can invoke this low degree of differentiability as a reason why these solutions should not be considered to violate cosmic censorship. On the other hand, it may be possible to impose the cut-off on the initial data surface in a less severe manner, i.e. match more smoothly to an asymptotically flat but non-empty initial data surface. If this were the case, the problem of the low degree of smoothness would be removed. There are of course further reasons to doubt the significance of these naked singularities, the most obvious of which is the highly specialised nature of these solutions, corresponding as they to do spherically symmetric, self-similar collapsing thick shells of matter with a light-like inner boundary and with no outgoing radiation.
Thus they fail the `seriousness' criterion which should be applied to would be counter-examples to cosmic censorship on the grounds of not supplying sufficiently general initial data. However this view must take account of the important role played by self-similarity in general relativity \cite{carr-coley}. Self-similar solutions arise as asymptotic end-states of various configurations in general relativity, including spatially homogeneous cosmologies and for some spherically symmetric situations. Self-similarity also plays a crucial role in the understanding of critical phenomena in gravitation \cite{gundlach}. There exists the possibility that realistic gravitational collapse can lead to an asymptotically self-similar state. If that is the case, one could then pick up regular initial data at a late stage of the collapse for which the assumption of self-similarity would be justified. In this situation, the solutions studied here would be of some relevance. 

A more serious objection is that, although we did so in the cause of generality, not specifying the matter distribution of the space-time means that we have left aside the question of whether or not the Einstein-matter field equations are well posed. (Indeed this applies not only to the matter model iteslf, but also to the non-smooth data with which we have worked.) This is an integral part of the cosmic censor conjecture (see e.g. Wald's textbook \cite{wald}.) So while the question of whether or not the matter chosen here is suitable for the study of cosmic censorship is open, it is likely that our results fit into an emerging picture wherein the hypothesis is verified - for small data, at least - for `good' matter models ({\em cf.} Christodoulou and Klainerman for vacuum \cite{ck}, Christodoulou for the scalar field \cite{christo} and Rein and Rendall for the Einstein-Vlasov system \cite{rr}), but violated for `poor' matter models ({\em cf.} Christodoulou \cite{christo1}, Newman \cite{newm}, Shapiro and Teukolsky \cite{st} and Joshi and Dwivedi \cite{jd} all, essentially, for dust). See \cite{rendall-matter,rendall-letter} for a detailed discussion on what comprises a good matter model.

Along with this issue, there are numerous important steps involved in  determining the significance of the results presented here. Among these is to analyse the stability of the Cauchy horizon to various perturbations. Another is to remove the assumption that $\psi^\prime(x)=0$ and so to include pure outgoing radiation. The main increase in the degree of difficulty of the problem would be that the energy conditions, which have played a central role here, change from being linear differential inequalities for one function to non-linear differential inequalities for two functions. This difficulty notwithstanding, it would be more surprising if the results here were not reproduced in the general case ($\psi^\prime(x)\neq 0$) than if they were: as Waugh and Lake have shown \cite{waugh-lake}, naked singularities in spherically symmetric homothetic collapse are identified with root sets of certain algebraic equations, just as is the case here. The space of solutions analogous to ${\cal {F}}$ here would be more complicated, but as the key identification marks of naked and censored singularities are respectively the existence and absence of roots of an algebraic equation, one would expect both subsets of solutions to be topologically stable. 

\section*{Acknowledgement}
I am grateful to an anonymous referee for constructive criticism.


\begin{thebibliography}{hkvf}
\bibitem[1]{waldrev}
R.~M. Wald, preprint gr-qc/9710068.

\bibitem[2]{ck}
D. Christodoulou and S. Klainerman, {\it The Global Nonlinear Stability of the Minkowski Space} (Princeton University Press, Princeton, N.J. 1993).

\bibitem[3]{klain}
S. Klainerman and F. Nicol\`{o}, Class. Quant. Grav. {\bf 16}, R73 (1999).

\bibitem[4]{rr}
G. Rein and A.~D. Rendall, Commun. Math. Phys. {\bf 150}, 561 (1992).

\bibitem[5]{christo}
D. Christodoulou, Ann. Math. {\bf 149}, 183 (1999).

\bibitem[6]{brady-kyoto}
P.~R. Brady, Prog. Theor. Phys. Suppl. {\bf 136}, 29 (1999).

\bibitem[7]{racz}
I. Racz, Class.Quant.Grav. {\bf 17}, 153 (2000).

\bibitem[8]{jj}
S. Jhingan and P.~S. Joshi, in {\it Internal structure of black
holes and spacetime singularities}, 
L.~M. Burko and A.
Ori (eds.), (IOP Publishing, Bristol, 1997).

\bibitem[9]{christo1}
D. Christodoulou, Ann. Math. {\bf 140}, 607 (1994).

\bibitem[10]{ori-piran}
A. Ori and T. Piran, Phys. Rev. D{\bf 42}, 1068 (1990).

\bibitem[11]{lake}
K. Lake and T. Zannias, Phys. Rev. D{\bf 42}, 3866 (1990).

\bibitem[12]{carr-coley}
B.~J. Carr and A.~A. Coley, Class. Quant. Grav. {\bf 16}, R31 (1999).

\bibitem[13]{gundlach}
C. Gundlach,
Living Reviews,\hfill\newline  {\tt http://www.livingreviews.org/Articles/Volume2/1999-4gundlach} (1999).

\bibitem[14]{gold-pir}
D. Goldwirth and T. Piran, Phys. Rev. D{\bf 36}, 3575 (1987).

\bibitem[15]{henrik}
R.~N. Henriksen and K. Patel, Gen. Rel. Grav. {\bf 23}, 527 (1991).

\bibitem[16]{brady}
P.~R. Brady, Phys. Rev. D{\bf 51}, 4168 (1995).

\bibitem[17]{waugh-lake}
B. Waugh and K. Lake, Phys. Rev. D{\bf 40}, 2137 (1989).

\bibitem[18]{HE}
S.~W. Hawking and G.~F.~R. Ellis, {\it The Large Scale Structure
of Space-Time} (Cambridge University Press, Cambridge, 1973).

\bibitem[19]{BI}
C. Barrabes and W. Israel, Phys. Rev. D{\bf 43}, 1129 (1991).

\bibitem[20]{nolan99}
B.~C. Nolan, Phys. Rev. D{\bf 60}, 024014 (1999).

\bibitem[21]{wald}
R.~M. Wald, {\it General Relativity} (University of Chicago Press, Chicago, 1984).

\bibitem[22]{newm}
R.~P.~A.~C. Newman, Class. Quant. Grav. {\bf 3}, 527 (1986).

\bibitem[23]{st}
S.~L. Shapiro and S.~A. Teukolsky, Phys. Rev. Lett. {\bf 66}, 994 (1991).

\bibitem[24]{jd} 
P.~S. Joshi and I.~H. Dwivedi, Phys. Rev. D{\bf 47}, 5357 (1993).

\bibitem[25]{rendall-matter}
A.~D. Rendall, in {\it Approaches to numerical relativity}, 
R. d'Inverno (ed), (Cambridge University Press, Cambridge, 1992).

\bibitem[26]{rendall-letter}
A.~D. Rendall, Class. Quant. Grav. {\bf 9}, L99 (1992).

\end{thebibliography}
\end{document}